\documentclass[12pt]{article}
\usepackage{epsfig}

\makeatletter
\def\vereq#1#2{\lower3pt\vbox{\baselineskip1.5pt \lineskip1.5pt
\ialign{$\m@th#1\hfill##\hfil$\crcr#2\crcr\sim\crcr}}}
\makeatother

\begin{document}

\newcommand{\mgrav}{m_{3/2}}
\newcommand{\LS}{\Lambda_{\rm SUSY}}
\newcommand{\Mp}{M_{\rm pl}}

\begin{titlepage}

\begin{flushright}
CU-TP-1027\\
LBNL-48726\\
UCB-PTH-01/32\\
astro-ph/0108172\\
\end{flushright}
\begin{centering}
\vspace{.2in}
{\Large {\bf Gravitino Warm Dark Matter\\
\vspace{.05in}with Entropy Production}}\\
\vspace{.4in}
Edward A. Baltz\\ \vspace{0.1in}
{\footnotesize\it
Columbia Astrophysics Laboratory, 550 W 120th St., Mail Code 5247,
New York, NY 10027\\
{\tt eabaltz@physics.columbia.edu}}
\vspace{0.3in}

Hitoshi Murayama\\ \vspace{0.1in}
{\footnotesize\it
Theory Group, Lawrence Berkeley National Laboratory and\\ Department of
Physics, University of California, Berkeley, CA 94720\\
{\tt murayama@lbl.gov}}
\vspace{1in}

{\bf Abstract} \\
\end{centering}
\vspace{0.2in}
\hspace{.2in} Gravitinos with a mass in the keV range are an interesting
candidate for warm dark matter.  Recent measurements of the matter density of
the universe and of cosmic structures at the dwarf galaxy scale rule out the
simplest gravitino models with thermal freeze--out.  We construct a model where
the decay of the messenger particles that transmit the supersymmetry breaking
to the observable sector generates the required entropy to dilute the gravitino
relic density by the required factor of a few to come in line with
observations.  The model is natural, and requires only that the coupling of the
messenger sector to the standard model be set so that the decay happens at the
appropriate time.

\end{titlepage}

\section{Introduction}

We revisit the possibility that gravitinos could be the dominant component of
nonrelativistic matter in the universe.  Especially interesting to us are
scenarios where the gravitino is very light, with a mass in the keV range.
Such a species would be ``warm'' dark matter.

This possibility is interesting for two reasons, which we will now
enumerate.  The first is that such a light gravitino arises naturally
in models of gauge--mediated supersymmetry breaking \cite{gmsb} (GMSB)
revived recently \cite{gmsbdiscuss}.  They are stable, and thus could
potentially pose problems for cosmology
\cite{gmsbcosmology,gmsbcosmology2,gmsbcosmology3}, which we will
address in light of new more accurate measurements of both the matter
density of the universe and on constraints on deviations from cold
dark matter.

Our second motivation is that recently cold dark matter has come under fire,
and warm dark matter may in fact be favored.  In particular, the predictions of
the cold dark matter theory indicate that galaxies such as the Milky Way should
have a large number of satellites, which are not observed.  Warm dark matter
could ameliorate this problem, as density fluctuations at the satellite scale
and smaller are suppressed \cite{Dolgov, BOT}.  In these models, the densities
of halo cores are lowered, large halos have many fewer subhalos, and the
subhalos that are present form in a top--down fashion, namely by fragmentation
of larger clumps.  This is contrary to the prediction of cold dark matter,
where all relevant structures form in a bottom--up fashion by mergers of
smaller structures \cite{BOT}.

As data have improved, both in terms of the matter density of the universe, and
in constraints on deviations from collisionless dark matter, the gravitino has
silently stopped being a viable dark matter candidate as its relic density is
too large in the simple thermal freeze-out scenario.  We rescue the gravitino
in GMSB models by invoking a natural model for entropy production, and thus
provide again a candidate for warm dark matter, with a mass in the range
$1.0-1.5$ keV.

\section{Gravitino mass and relic density}

In GMSB models, the gravitino mass $\mgrav$ is related to the SUSY breaking
scale $\LS$ as follows,
\begin{equation}
\mgrav=\frac{1}{\sqrt{3}}\left(\frac{\LS^2}{\Mp}\right)=0.237\;\;{\rm keV}\;
\left(\frac{\LS}{\rm PeV}\right)^2,
\end{equation}
where $\Mp$ is the reduced Planck mass.  We will be concerned with
gravitino masses around 1 keV, thus SUSY breaking scales of 2 PeV,
which are natural in GMSB models.\footnote{In this class of models,
  so-called messenger particles acquire a mass $M$ and a
  supersymmetry-breaking scalar mass term $F$.  The superparticles in
  the SUSY Standard Model acquire soft masses via gauge interactions
  at the 2--loop level $m^2_{\rm soft} = \sum_a 2 C_a
  \left(\frac{g_a^2}{16\pi^2}\right)^2\; \left(\frac{F}{M}\right)^2$,
  where $g_a$ are gauge coupling constants and $C_a$ are quadratic
  Casimir.  In the original model of gauge mediation, $\sqrt{F} \sim
  M$ is suppressed relative to the primordial supersymmetry breaking
  scale $\LS$ by a loop-factor, and the gravitino mass is typically
  heavier than 50~keV \cite{gmsbcosmology2}.  But in models of direct
  gauge mediation \cite{direct}, $\sqrt{F} \sim \LS$ with no loop
  suppression.  In particular, the model in Ref.~\cite{INTY} achieves
  $m_{3/2} \sim 1$~keV naturally.  We thus can have standard
  weak--scale supersymmetry in the rest of the superparticle spectrum,
  with the caveat being that the gravitino is the LSP and all other
  superparticles decay to gravitinos and Standard Model particles.}

Gravitinos in this scenario interact fairly strongly, and thus freeze out quite
late, at a temperature of order the weak scale or even colder.  As they are of
course highly relativistic, the computation of their relic density is
relatively simple: all that needs to be done is to compute the effective
degrees of freedom $g_*$ at freeze--out.  Their relic density is then obtained,
\begin{equation}
\Omega_{3/2}h^2=1.14\left(\frac{g_*}{100}\right)^{-1}\left(\frac{\mgrav}{\rm
keV}\right).
\end{equation}
The value of $g_*$ at gravitino freeze--out has been determined to be in the
range 90-140 for a wide range of parameters \cite{RD}.

Finally we consider astrophysical constraints on the mass of the relic particle
making up the dark matter.  Particles in the eV range that undergo thermal
freeze--out are ruled out as ``hot'' dark matter.  The transition to ``cold''
dark matter occurs for thermal relics in the keV range.  In fact, there is a
lower bound on the mass of a warm relic \cite{BOT,LyA},
\begin{equation}
\mgrav>{0.75-1.0\;{\rm keV}}\longrightarrow\Omega_{3/2}h^2>0.6,
\end{equation}
based on the requirement that the small scale fluctuations are not in conflict
with e.g. the Ly$\alpha$ forest.  This mass bound is valid for any {\em
thermal} relic, but is modified for relics whose distribution functions deviate
from equilibrium \cite{steen}.

We compare this bound on the relic density of gravitinos with current data on
the density of the dark matter, which favors $\Omega_{\rm DM}h^2\approx0.15$
\cite{cmbdata}, and in any circumstances $\Omega_{\rm DM}h^2<0.3$, indicating a
serious discrepancy.  One way to alleviate this is to allow for entropy
production after gravitino freeze--out, which dilutes the gravitino relic
density to acceptable levels.  We discuss this possibility in the next section.

\section{Entropy production: simple model}

The production of entropy is required to dilute the relic density of
gravitinos.  As current data favor $\Omega_{\rm DM}h^2\approx0.15$, and in any
event $\Omega_{\rm DM}h^2<0.3$, we will consider a comoving entropy density
increase of a factor of several.

In GMSB models there is a messenger sector around the SUSY breaking scale,
which for keV gravitinos is of order 2 PeV.  As our entropy production
mechanism we will consider the out-of-equilibrium decays of the messengers via
small couplings to the Standard Model.  Assuming stability of the messengers,
their relic density has been calculated \cite{hh}, and we simply quote the
results here,
\begin{equation}
\Omega_M h^2\approx10^5\left(\frac{M}{\rm PeV}\right)^2\longrightarrow
Y_M=\frac{\Omega_M\rho_c}{s_0M}
\approx3.65\times10^{-10}\left(\frac{M}{\rm PeV}\right),
\end{equation}
where $M$ is the mass of the messenger particle, $Y_M$ is the frozen--out value
of $n/s$, the ratio of number and entropy densities for the messenger
particles, $s_0$ is the entropy density today, and $\rho_c$ is the critical
density today.

In this section we will assume that the messengers decay instantaneously when
the temperature of the thermal bath is $T_D$, and their Standard Model decay
products thermalize instantaneously also.  This approximation requires only the
conservation of energy.  Before decay, the energy density is
\begin{equation}
\rho=\frac{\pi^2}{30}\,g_*T^4+\frac{2\pi^2}{45}\,g_*T^3MY_M.
\end{equation}
The first term is the radiation density and the second is the matter density of
messengers.  This quantity is simply equated to the radiation density
afterwords, with a higher temperature $T'$, and no matter, giving the
temperature increase.  This is then simply related to the fractional increase
in entropy density $s'/s$.
\begin{equation}
\left(\frac{T'}{T}\right)^4=\left(\frac{s'}{s}\right)^{4/3}=
1+\frac{4}{3}\,Y_M\left(\frac{M}{T}\right)=
1+4.87\left(\frac{M}{\rm 10\;PeV}\right)^2
\left(\frac{T_D}{\rm 10\;MeV}\right)^{-1}.
\end{equation}
Here we gave the temperature in MeV to be suggestive.  The temperature at decay
must be larger than this so as not to disrupt the nucleosynthesis era.  To be
safe, $T_D>$10 MeV is warranted.

To be specific, consider a gravitino with mass 1 keV, and thus
$\Omega_{3/2}h^2\approx 1$.  Assuming we want $\Omega h^2=0.15$, we must dilute
by a factor of $s'/s=6$.  For this model we find that the temperature increases
by a factor of $T'/T\approx 1.8$ during the decay.  We now find a simple
relation between the mass of the messenger and the temperature at which it
decays,
\begin{equation}
\left(\frac{M}{\rm 10\;PeV}\right)^2\approx
2\left(\frac{T_D}{\rm 10\;MeV}\right).
\end{equation}
All that remains to be done in this model is to adjust the coupling between the
messengers and the Standard Model particles to enforce this constraint on the
lifetime $\tau$ of the messengers,
\begin{equation}
\tau\equiv\frac{1}{\Gamma}=\frac{1}{2H}\equiv\frac{\Mp}{T_D^2}
\sqrt{\frac{45}{2\pi^2g_*}}
\longrightarrow\left(\frac{T_D}{\rm 10\;MeV}\right)^2\approx1.11\times10^{29}\,
\frac{\Gamma}{10\;{\rm PeV}}.
\end{equation}
Note that here $g_*$ is the effective number of degrees of freedom at the
decay, which for temperatures between 1 and 100 MeV is $g_*=43/4$.  Therefore,
the coupling of messenger particles to the standard model particles must be
extremely suppressed.  This is actually a desirable feature so as to not spoil
the flavor-independence of the gauge-mediated supersymmetry breaking
\cite{DNS}.

If the messengers decay directly to light particles, we might expect
$\Gamma\sim g^2M$, which would mean there is an extreme fine tuning of the
coupling $g$.  However, if the interaction is mediated by heavy particles
of masses $M_X$, the lifetime is
\begin{equation}
\Gamma\approx\frac{g^4}{192\pi^3}\frac{M^5}{M_X^4}
\rightarrow
\left(\frac{T_D}{\rm 10\;MeV}\right)^2\approx
18.6\left(\frac{g^2}{0.01}\right)^2\left(\frac{M}{10\;{\rm
PeV}}\right)^5\left(\frac{M_X}{10^{12}\;{\rm GeV}}\right)^{-4}.
\end{equation}
Thus, a heavy particle of a mass of $10^{12}$ GeV coupling the messenger sector
to the standard model gives roughly the right lifetime.  This is somewhat
lower than the GUT scale, but still far above the messenger scale, so these
interactions should not invalidate the relic density prediction.

\section{Entropy production: full evolution}

In the previous section we assumed that the messengers decay instantaneously.
This approach allows us to study entropy production without lengthy
calculations.  The results from a more accurate calculation are similar, as we
will show \cite{KT}.  In reality of course, the decay is exponential, and we
should follow the evolution of matter and radiation through the decay process
to accurately determine the increase in entropy, and the relative change in
gravitino temperature.  In this section we will do just that.  We find a
similar increase in entropy as for the simple calculation, though we also show
that the temperature never increases, it merely decreases more slowly.

We now write down the equations for the evolution of the radiation and matter
densities, and the expansion scale factor.  We assume that the universe is
flat, with no cosmological constant, as is appropriate for early times.  The
equations are
\begin{eqnarray}
\dot{\rho}_M+3H\rho_M&=&-\Gamma\rho_M,\label{eq:matter}\\
S^{1/3}\dot{S}&=&\left(\frac{2\pi^2}{45}\,g_*\right)^{1/3}\Gamma\rho_MR^4,\label{eq:S}\\
H^2\equiv\left(\frac{1}{R}\frac{dR}{dt}\right)^2&=&
\frac{1}{3\Mp^2}\left(\rho_M+\rho_R\right),\label{eq:scale}
\end{eqnarray}
where $S=sR^3$ is the comoving entropy density.  The first equation is solved
trivially,
\begin{equation}
\rho_M(t)=\rho_M(t_i)\left(\frac{R(t)}{R(t_i)}\right)^{-3}e^{-\Gamma(t-t_i)},
\end{equation}
and we solve the remaining two equations numerically.  We recover the entropy
increase from the solution,
\begin{equation}
\frac{s'}{s}=\frac{S(t_f)}{S(t_i)}.
\end{equation}
The effective temperature decrease of the gravitino is also easily recovered as
\begin{equation}
\frac{T_R(t_f)/T_R(t_i)}{T_{3/2}(t_f)/T_{3/2}(t_i)}
=\left(\frac{s'}{s}\right)^{1/3},
\end{equation}
in other words the radiation temperature decreases less than the gravitino
temperature.  This indicates that for a given mass, the gravitino behaves more
like cold dark matter than expected.

Considering the results of the previous section, we choose two fiducial cases
to run numerically.  We use decay temperatures of 10 and 100 MeV, requiring
$M\approx 14,45$ PeV respectively, and we set the decay times as appropriate.
Evolving Eqs.~(\ref{eq:matter}--\ref{eq:scale}) for these parameters, we arrive
at Fig.~1.  We start the evolution at a temperature of 10 GeV, and take
$g_*(T)$ for entropy from Ref.~\cite{GG}.  The onset of matter domination and
the entropy--producing decay are both clear.  By a temperature of a few MeV,
the decay is all but complete, and nucleosynthesis can proceed as normal.  We
note that in principle, an epoch of matter domination such as is present in our
model affects the evolution of the density fluctuations responsible for
structure formation.  In our case, the effects occur at wavelengths far too
short to be observed in the cosmic microwave background or in large scale
structure.

\begin{figure}
\epsfig{file=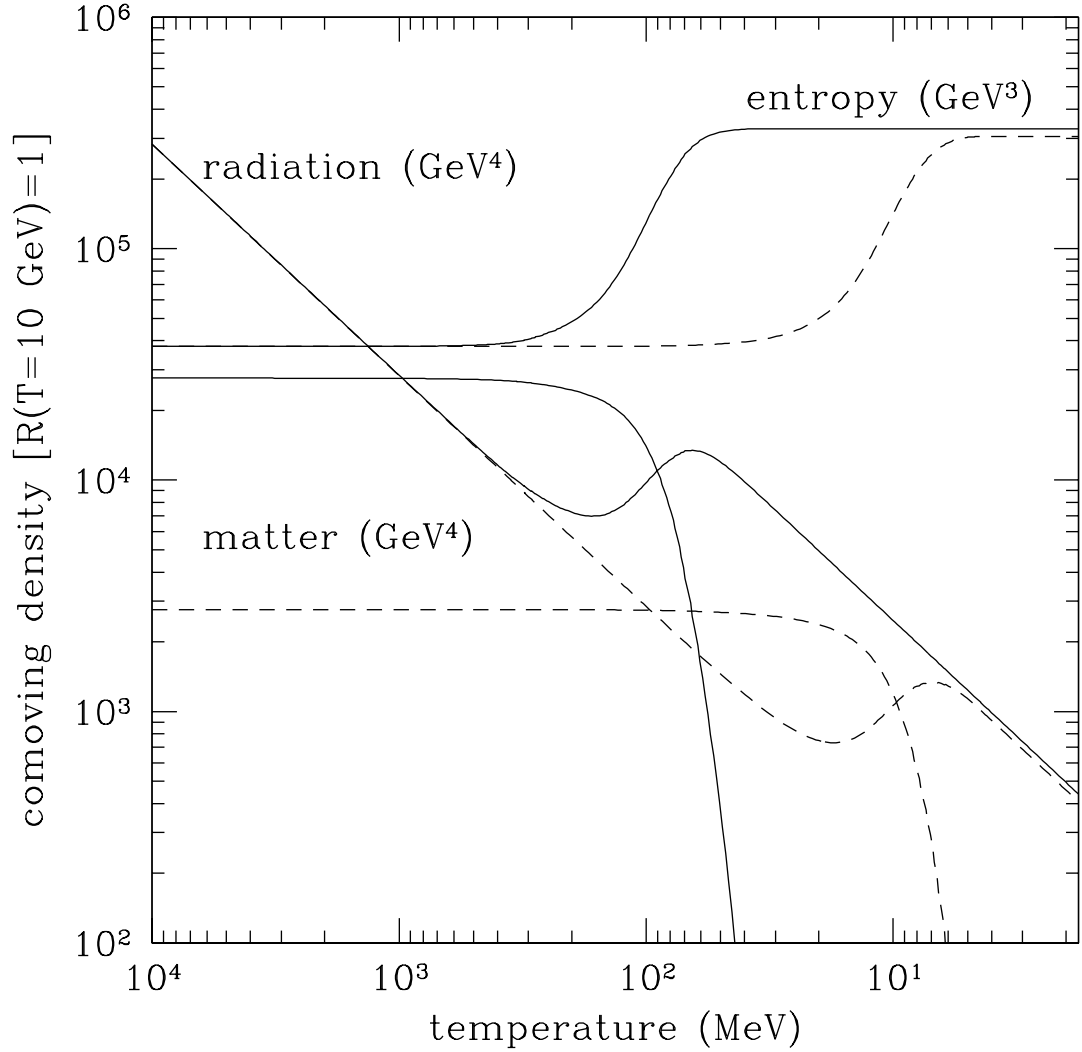,width=6in}
\caption{Evolution of entropy density.  Here we show the decay of messengers at
temperatures of 100 MeV (solid curve) and 10 MeV (dashed curve), producing an
entropy increase of about a factor of eight.  The onset of matter domination at
a temperature of about 1 GeV (100 MeV) is clearly seen, and the decay at 100
MeV (10 MeV) is also clear.}
\end{figure}

\section{Discussion}

We have constructed a model for gravitino warm dark matter that has the
appropriate relic density.  Entropy is produced by the decay of the messenger
particles.  The lifetime of the messengers is of the correct order if the decay
happens through heavy particles of $10^{12}$~GeV or so.

For the keV gravitino we consider, the temperature, and thus the momenta, are
decreased relative to the radiation by a factor of approximately two.  This
modifies the free--streaming scale $R_{\rm fs}$ of the gravitinos, and thus the
effect on structure formation.  This length scale is proportional to the
velocity of the particles at the epoch of matter--radiation equality, and thus
$R_{\rm fs}\propto T_{3/2}/\mgrav$, normally given by \cite{bs,bbks}
\begin{equation}
R_{\rm fs}\approx
0.2\left(\frac{g_*}{100}\right)^{-1/3}\left(\frac{\mgrav}{\rm keV}\right)^{-1}
{\rm Mpc}\approx
0.2\left(\Omega_{3/2}h^2\right)^{1/3}
\left(\frac{\mgrav}{\rm keV}\right)^{-4/3}\;{\rm Mpc},
\end{equation}
where in the second relation the expression for the relic density is
substituted for mass.  In our case the relic density is different, and we must
be careful in replacing the expression.  With entropy production, the
gravitinos are colder at matter--radiation equality than a similar constituent
would be without entropy production.  This free--streaming scale comes about at
the epoch of matter--radiation equality, where the temperature is a few eV.
The gravitinos at this time are non-relativistic, so we can simply decrease
their velocity by the decrease in temperature.  We thus find that roughly
speaking,
\begin{equation}
R_{\rm fs}\approx0.2\left(\frac{g_*(s'/s)}{100}\right)^{-1/3}\left(\frac{\mgrav}
{\rm keV}\right)^{-1}{\rm Mpc}
\approx0.2
\left(\Omega_{3/2}h^2\right)^{1/3}
\left(\frac{\mgrav}{\rm keV}\right)^{-4/3}\;{\rm Mpc}.
\end{equation}
This expression is identical to the previous one.  If we use the quantities
$\mgrav$ and $\Omega_{3/2}h^2$ only, the expression for the free--streaming
scale is unaffected by entropy production, at least in this approximation.

A more accurate approach accounts for free--streaming prior to
matter--radiation equality \cite{BOT,steen}.  We simply quote the results here.
\begin{equation}
R_{\rm fs}\approx 0.65\left(\Omega_mh^2\right)^{0.15}
\left(\frac{\mgrav}{\rm keV}\right)^{-1.15}\;{\rm Mpc}.
\end{equation}

As the current bounds on the mass of a warm dark matter particle are in the
$0.75-1.0$ keV range, we consider a gravitino slightly more massive than this,
in the range $1.0-1.5$ keV.  The beneficial qualities of the warm dark matter
scenario for the most part persist for particles in this mass range \cite{BOT}.
The entropy production required is then
\begin{equation}
\frac{s'}{s}=7.6\left(\frac{g_*}{100}\right)^{-1}
\left(\frac{\Omega_mh^2}{0.15}\right)^{-1}\frac{\mgrav}{\rm keV}.
\end{equation}
We have shown that entropy production of this order is not a serious
difficulty, thus the gravitino is again a viable candidate for warm dark
matter.

\section*{Acknowledgements}

We gratefully acknowledge the hospitality of the organizers of
PASCOS'01, where this research was begun.  HM was supported in part by
the U.S.~Department of Energy under Contract DE-AC03-76SF00098, and in
part by the National Science Foundation under grant PHY-95-14797.

\end{document}